\documentclass[10pt]{article}
\usepackage[letterpaper,margin=0.72in,left=2.05in]{geometry}
\usepackage{graphicx,amsmath,booktabs,xcolor,hyperref,fancyhdr,tikz,eso-pic,microtype}
\usepackage{newtxtext,newtxmath}
\definecolor{jossblue}{RGB}{0,75,200}
\hypersetup{colorlinks=true,linkcolor=jossblue,citecolor=jossblue,urlcolor=jossblue}
\begin{document}
{\LARGE\bfseries exa-PD: A scalable high-performance workflow for\\multi-element phase diagram construction}\par
\vspace{8pt}
{\large\bfseries Zhuo Ye$^{1}$, Feng Zhang$^{1}$, Maxim Moraru$^{2}$, Weiyi Xia$^{1}$, Ying Wai Li$^{2}$, Yongxin Yao$^{1}$, and Cai-Zhuang Wang$^{1}$}\par
\smallskip
$^{1}$ Ames National Laboratory, U.S. Department of Energy, Ames, Iowa 50011, USA\\
$^{2}$ Los Alamos National Laboratory, Los Alamos, NM 87545, United States of America

\section*{Summary}
\begin{figure}[ht]\centering\includegraphics[width=0.58\linewidth]{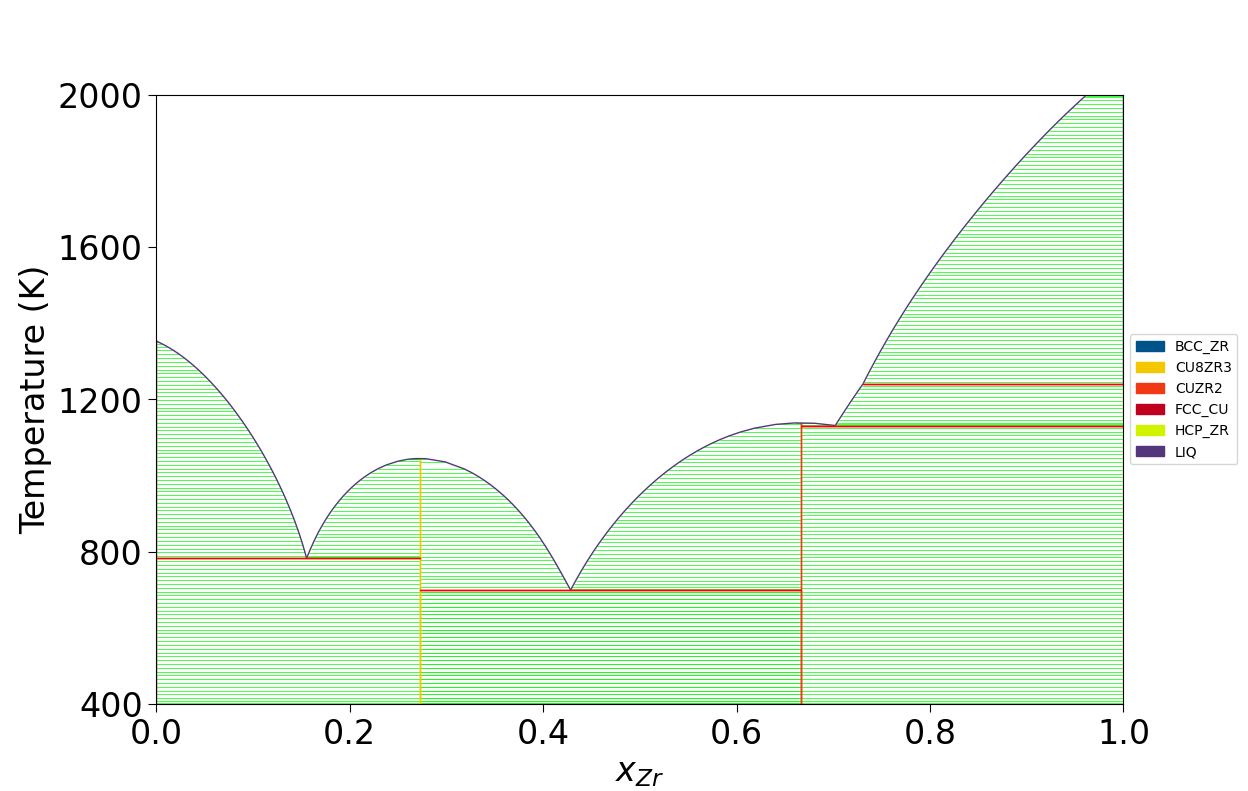}\caption{Phase diagram of the Cu--Zr system predicted by exa-PD using an EAM-FS potential.}\label{fig:pd}\end{figure}

Exa-PD is a highly parallelizable workflow designed for the construction of multi-element phase diagrams (PDs) \cite{exa-PD}. It uses standard sampling techniques---molecular dynamics (MD) and Monte Carlo (MC)---as implemented in the LAMMPS package \cite{thompson2022}, to simultaneously sample multiple phases over a fine temperature--composition mesh for free-energy calculations. Parsl \cite{babuji2019} serves as the global workflow engine, coordinating large ensembles of MD and MC tasks to achieve massive parallelization with strong scalability. The resulting free energies of liquid and solid phases are then fed to CALPHAD modeling via the PyCalphad package \cite{otis2017} to construct multi-element PDs, as illustrated in Figure~\ref{fig:pd}.

By leveraging Parsl, a parallel programming library for Python, exa-PD enables scalable execution of large ensembles of MD and MC simulations with internal task dependencies across heterogeneous resources. The framework supports efficient scaling from a single workstation to multi-node supercomputers and exhibits near-linear scaling up to 32 GPUs and good scaling up to 32 CPUs through dynamic task distribution, as shown by the benchmark results in Ref.~\cite{zhang2025} and in Figure~\ref{fig:scaling} of this work.

\section*{Statement of Need}
Computational materials discovery has progressed rapidly with advances in computing and AI/ML techniques. However, experimental validation remains limited, largely due to insufficient knowledge of viable synthesis pathways. Reliable multi-element phase diagrams are therefore essential for resolving thermodynamic phase competition under synthesis conditions and for predicting synthesizability. Constructing these phase diagrams computationally requires highly accurate free-energy calculations.

Traditional workflows \cite{menon2021} have established robust and efficient methods for computing free energies. They mainly focus on the accuracy and efficiency of individual free-energy calculations. In contrast, exa-PD addresses a complementary challenge: the scalable coordination of large ensembles of MD/MC jobs to get the free energies required for multi-element phase diagram construction. The implementation of a global controller using Parsl ensures massive parallelization with strong scalability, efficiently managing MD/MC jobs to handle resource-intensive calculations. By abstracting task execution into a flexible dependency graph, Parsl enables a data-driven execution model in which tasks are triggered as soon as their required inputs become available. While many MD tasks leverage GPU acceleration, certain essential features remain restricted to CPU execution, making heterogeneous CPU/GPU resources necessary. Parsl accommodates such heterogeneous execution across multiple node types, thereby enabling high-throughput management of MD simulations for free-energy calculations.

\section*{Workflow Overview}
\begin{figure}[ht]\centering\includegraphics[width=0.72\linewidth]{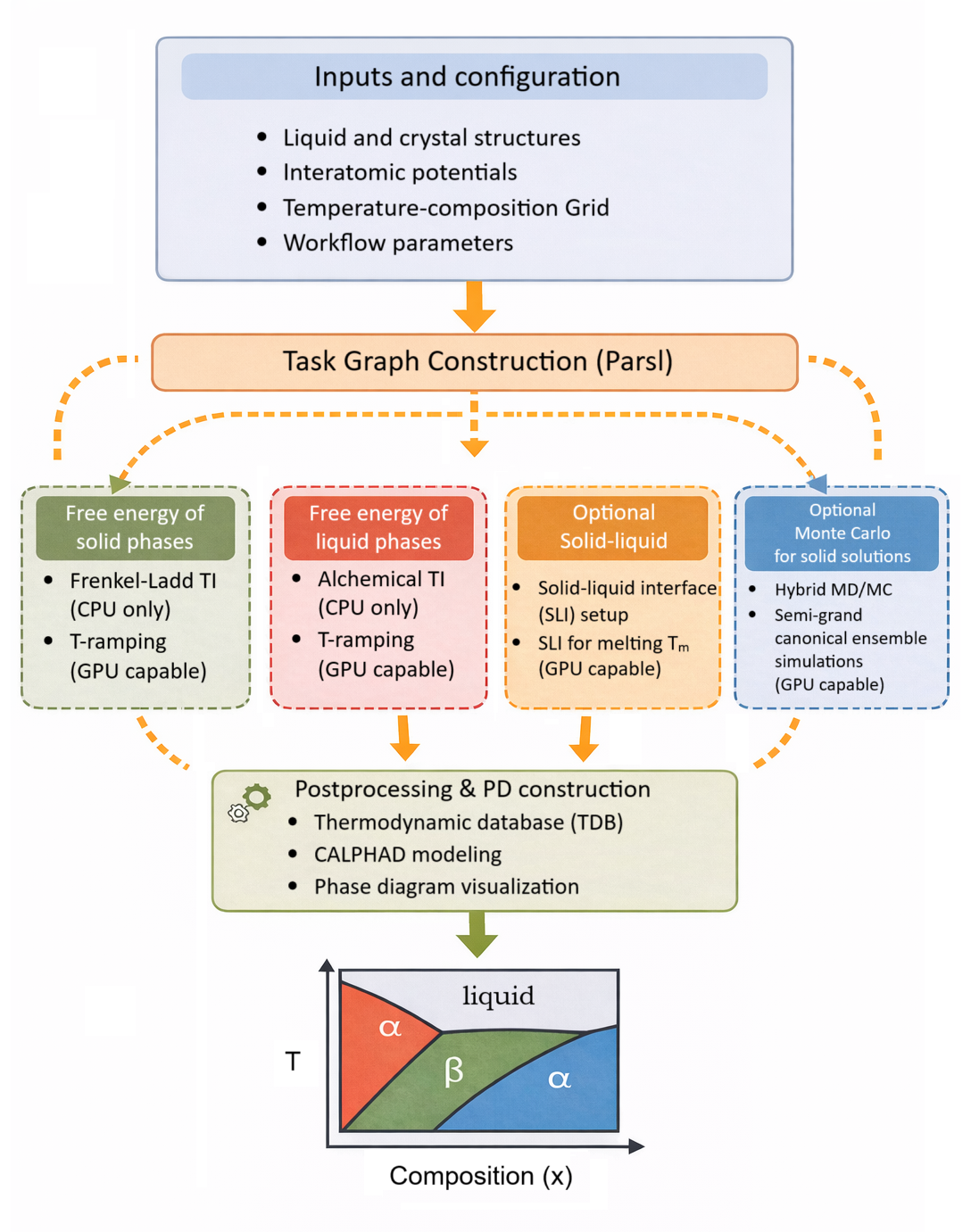}\caption{Schematic flowchart of the exa-PD workflow.}\label{fig:workflow}\end{figure}

Figure~\ref{fig:workflow} gives a schematic flowchart of the exa-PD workflow, outlining the required and optional MD/MC jobs for constructing a phase diagram, with the internal dependencies among the MD/MC jobs indicated by orange arrows. In principle, one needs the absolute free energy of solid and liquid phases to generate a phase diagram. The absolute free energy can be obtained by using a reference system, of which the free energy can be analytically derived, and calculating the free energy difference between the target and reference systems with thermodynamic integration (TI). The first module of the workflow calculates the free energy of solid phases (line compounds) by using a reference system such as an Einstein crystal at a certain temperature and estimating the Frenkel--Ladd TI. Then it ramps up or ramps down the temperature to obtain the free energies at other temperatures by integrating the Gibbs--Helmholtz equation. The second module calculates the free energy of liquids. Taking a binary system $A_{1-x}B_x$ as an example, the free energy of pure liquid $A$ is first obtained using the Uhlenbeck--Ford model (UFM) as the reference system, then the alchemical TI is used to obtain the free energy of other compositions $A_{1-x}B_x$. The third module is optional and determines the melting temperature of a certain solid phase using the solid--liquid coexistence (SLC) technique. SLC simulations are helpful to validate the free-energy results.

The execution model of exa-PD relies on Parsl's dynamic task distribution. Each module includes multiple asynchronous tasks that Parsl distributes across available computing resources. While many MD tasks benefit from GPU acceleration, some essential calculations are currently limited to CPU execution. For example, Frenkel--Ladd and alchemical TI calculations rely on LAMMPS features that are not supported by existing GPU or KOKKOS backends. However, when using the pre-compiled LAMMPS executable in the DeepMD package to implement DeepMD neural network potentials, all calculations presented in Figure~\ref{fig:workflow} can be executed efficiently on GPUs, enabling flexible utilization of heterogeneous CPU and GPU resources.

\begin{figure}[ht]\centering\includegraphics[width=0.86\linewidth]{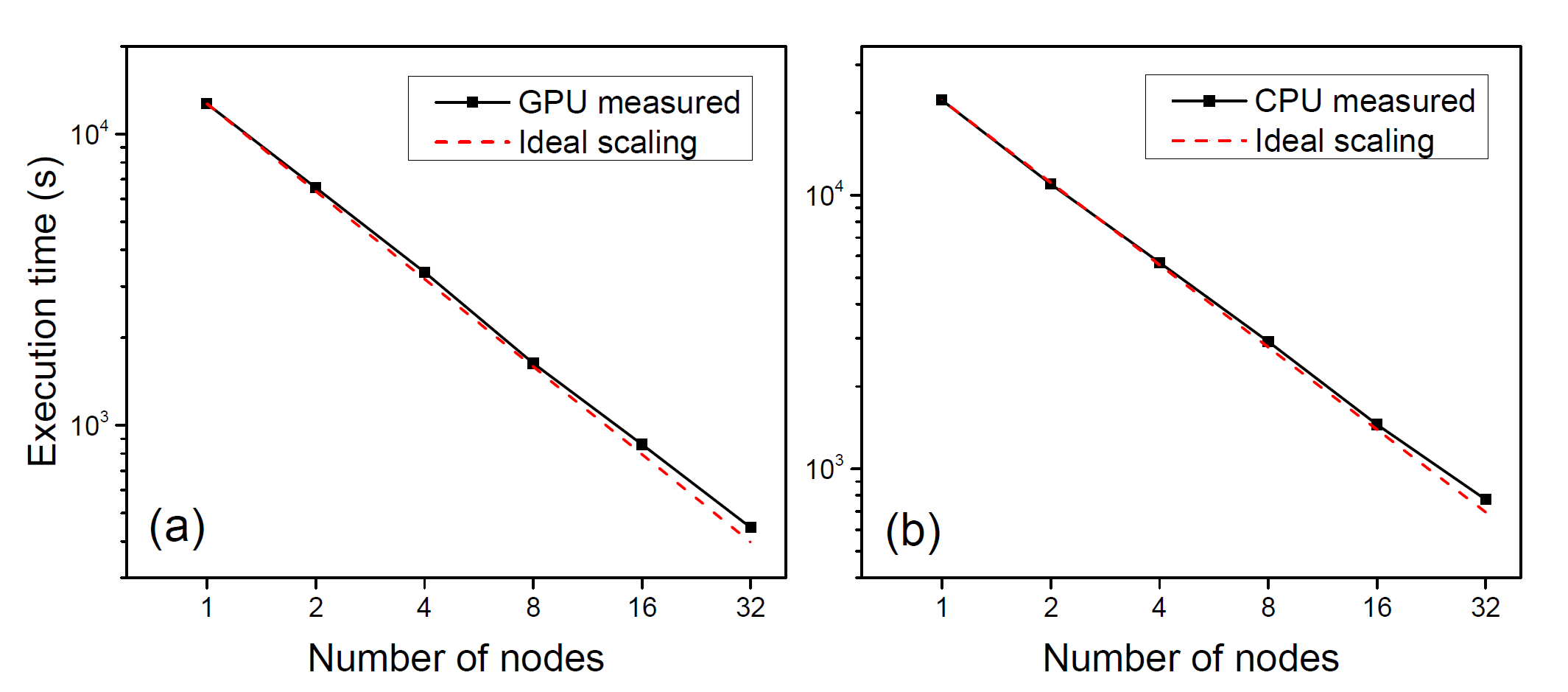}\caption{Strong scaling of exa-PD workflow on NERSC's Perlmutter supercomputer. Wall-clock times are shown for both GPU and CPU nodes.}\label{fig:scaling}\end{figure}

Figure~\ref{fig:scaling} summarizes the strong scaling results for the Cu--Zr system, evaluated on both GPU and CPU nodes. The benchmark tests are performed on Perlmutter at the National Energy Research Scientific Computing Center (NERSC). The execution time is plotted as a function of the number of compute nodes, along with the ideal linear scaling behavior for reference. Both GPU and CPU results closely follow the ideal scaling trend up to 32 nodes, achieving parallel efficiencies of approximately 89\% and 90\%, respectively.

In addition, exa-PD leverages Parsl to manage the complex dependency structure among hundreds of MD tasks. For example, the solid phase must first be equilibrated to determine the equilibrium volume and mean square displacement (MSD) for each species, in order to initialize the Frenkel--Ladd TI. Such task dependencies are expressed and enforced by the Parsl controller through its futures-based execution model.

After completion of all MD tasks, the results are postprocessed using the \texttt{run\_process.py} script. For each solid phase, the script generates a two-column dataset of Gibbs free energy $G$ as a function of temperature $T$. For the liquid phase, it produces a multi-column dataset of $G(T,x)$, with each column corresponding to a distinct composition. The postprocessing step also generates a thermodynamic database in TDB format that includes entries for all computed solid and liquid phases. Additionally, a sample script, \texttt{plot\_PD.py}, is provided to visualize the Cu--Zr phase diagram using PyCalphad based on the generated TDB file.

\section*{Initial Crystal Structures}
Exa-PD requires crystal structures for solid phases as input to free-energy calculations. It accepts unit-cell structures provided in popular formats, including the Crystallographic Information File (CIF) format and Vienna Ab initio Simulation Package (VASP) POSCAR files \cite{kresse1996}. In addition, exa-PD supports standard LAMMPS input files with the \texttt{.lammps} extension.

\section*{AI usage disclosure}
Generative AI tools were used to assist with language polishing and figure layout refinement. All technical content was reviewed and verified by the authors.

\section*{Acknowledgements}
This work was supported by the U.S. Department of Energy (DOE), Office of Science, Basic Energy Sciences, Materials Science and Engineering Division through the Computational Material Science Center program. Ames National Laboratory is operated for the U.S. DOE by Iowa State University under contract DE-AC02-07CH11358. Los Alamos National Laboratory is operated by Triad National Security, LLC, for the National Nuclear Security Administration of the U.S. Department of Energy under Contract No. 89233218CNA000001. This research used resources provided by NERSC under Contract No. DE-AC02-05CH11231 and resources provided by the Los Alamos National Laboratory Institutional Computing Program.

\end{document}